\documentclass[twocolumn,amssymb,fleqn, twocolumns]{revtex4} 
\usepackage{epsfig,amssymb,amsmath,graphicx,subfigure,hyperref  }  
\usepackage{color}

\usepackage{multirow} 
\usepackage{tabularx}

\begin{document}

\title{Stress-induced ordering of two-dimensional packings of elastic spheres}
\author{Zhenwei Yao}
\email{zyao@sjtu.edu.cn}
\affiliation{School of Physics and Astronomy, and Institute of Natural
Sciences, Shanghai Jiao Tong University, Shanghai 200240, China}
\begin{abstract}
  Packing of particles in confined environments is a common problem in multiple
  fields. Here, based on the two-dimensional Hertzian particle model, we study
  the packing of deformable spherical particles under compression, and reveal the crucial
  role of stress as an ordering field in regulating particle arrangement.
  Specifically, under increasing compression, the squeezed particles
  spontaneously organize into regular packings in the sequence of triangular and
  square lattices, pentagonal tessellation, and the reentrant triangular
  lattice. The rich ordered patterns and complex structures revealed in this
  work suggest a fruitful organizational strategy based on the interplay of
  external stress and intrinsic elastic instability of particle arrays. 
\end{abstract}

\maketitle

\section{Introduction}

Achieving order in the packing of elementary constituents in confined
environments is an important problem arising in the study of simple
liquids~\cite{scott1964random, finney1970random, hansen1990theory}, amorphous
solids~\cite{zallen,weeks2015amorphous,philippe2018glass}, space
tessellation~\cite{wang1961proving, bowers1997regular, Okabe2009,
palatinus2010circle} and especially in extensive soft matter systems such as the
packing of granular materials~\cite{rintoul1996metastability, makse2000packing}
and the self-assembly of colloids~\cite{dinsmore2002colloidosomes,
ershov2013capillarity}, filaments~\cite{Cui2010,bruss2012non}, and other
supramolecular structures~\cite{palmer2007supramolecular, xu2019ordered}.
To achieve ordered packing, several strategies have been proposed based on patterned
substrates~\cite{ye2001self,hellstrom2013epitaxial}, designed particle-particle
interaction~\cite{ershov2013capillarity, meng2014elastic, liu2018capillary},
chemical decoration of particles~\cite{knorowski2012dynamics,
li2013topological}, and
electromagnetic fields~\cite{biswal2004rotational, tierno2008controlled}. The
resulting ordered patterns lay the foundation for the fabrication of new nano-
and microstructured materials and provide broad possibilities of
functionalization~\cite{zhang2010colloidal,li2011colloidal,wang2012colloids}.
Especially, ordered two-dimensional (2D) arrays of particles on the surface lead to unique and
intriguing properties~\cite{dinsmore2002colloidosomes, Bausch2003e,
wan2015effects, brojan2015wrinkling}.  Recent studies on
the phase behavior of 2D soft particle systems show the crucial role of volume
fraction on ordered packing~\cite{pamies2009phase, miller2011two,
zu2017forming}. These studies inspire us to explore the role of stress in the
packing of geometrically confined deformable particles. The imposed stress
naturally reduces the free space available to particles, and it also introduces
subtle kinetic effects on the microscopic elastic behavior of particles. These
combined effects of stress may lead to an elasticity-based organizational
principle for particle packing.

The goal of this work is to explore the role of stress as an ordering field in
regulating particle arrangement. Our model consists
of a collection of frictionless, spherical Hertzian particles confined in a
two-dimensional box of tunable size.  The soft particle model enables the
variation of volume fraction beyond the hard-sphere limit, and soft particles
also serve as artificial atoms to reveal fundamental processes in materials
design across many length scales~\cite{landau99a, pamies2009phase}. This model
system allows us to clarify a series of questions: How does order
arise from randomness under imposed stress? What kinds of order could be
realized in the system of elastic particles? What is the microscopic behavior of
defects in response to the external stress? To address these inquiries, we
perform numerical experiment that allows us to track the detailed microscopic
process in the movement of particles. The data are subject to the combined
geometric, statistical and force analysis for extracting the essential physics.
Here, we emphasize that in this work we focus on 2D packing. The
three-dimensional (3D) packing
problem is fundamentally different from the 2D case due to the intrinsic
geometric frustrations~\cite{nelson1989polytetrahedral, atkinson2014existence}.

In this work, we reveal the scenario of how the collection of deformable
particles adapt their relative positions to the squeezed space under
compression. It is found that the highly heterogeneous local stress eliminates
the randomness in the initial particle configuration, and guides the particles
to form regular crystal lattices and quasicrystal structures. Specifically, the
ordering of loosely packed particles is dominated by the entropic effect, and
that of the squeezed particles under larger compression is determined by the
minimization of elastic energy. We highlight the crucial role of the emergent
grain boundary structure as an energy absorber to stabilize the triangular
lattice, and the formation of square lattice resulting from a series of local
elastic instability events. Under larger compression, we reveal the tessellation
of deformable pentagons and the reentrant behavior. We also discuss
the issue of reversibility and the kinetic effect of stress in facilitating
ordered packing. This work illustrates the role of stress as an ordering field
that may be exploited in various contexts of soft matter packing.

\section{Model and method}

\begin{figure}[th]  
\centering 
\includegraphics[width=3in]{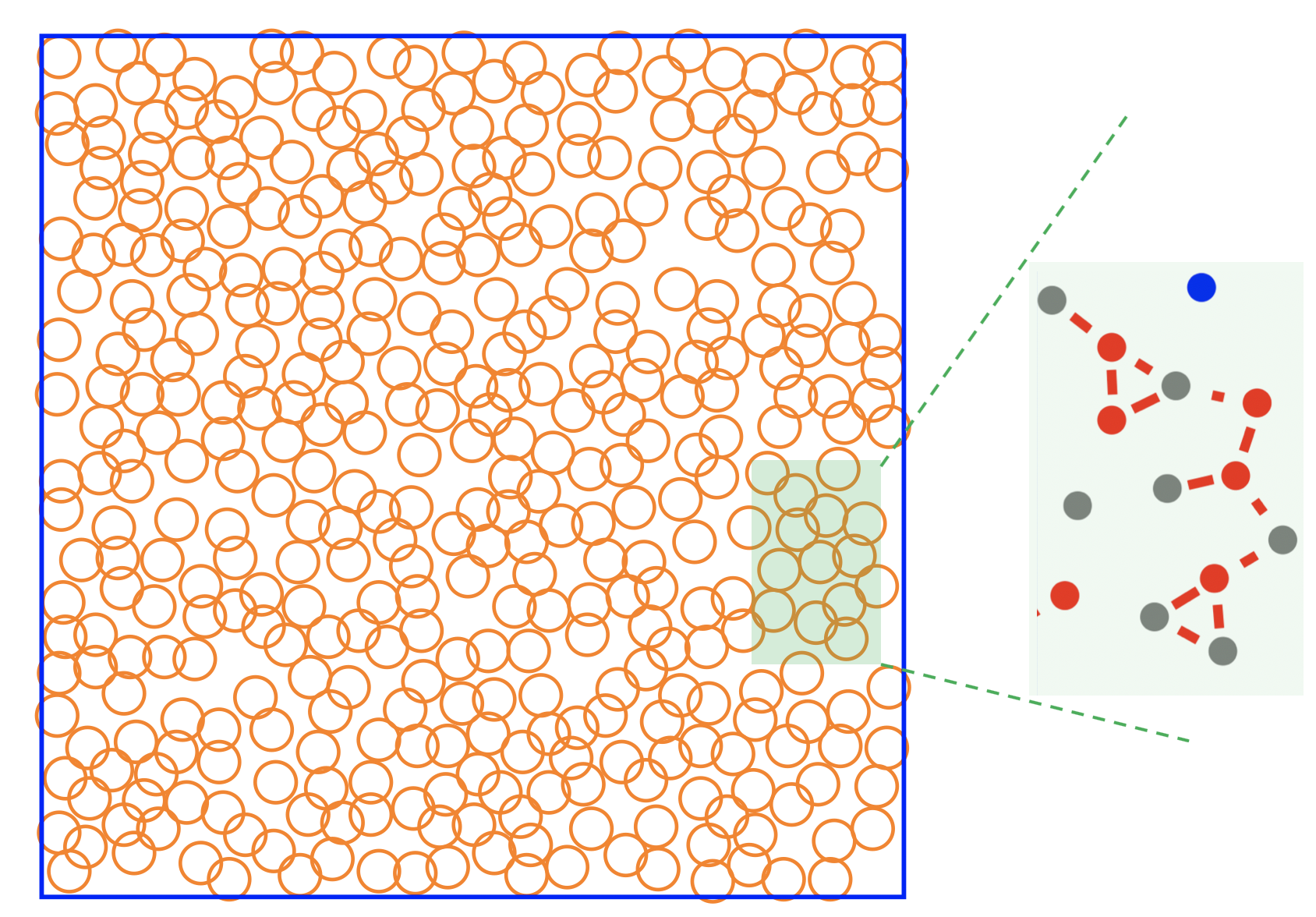}
  \caption{Schematic plot of the model system. The circles represent elastic
  particles of uniform size. The initial state is generated by random packing of
  particles. In the zoomed-in plot, the colored dots represent
  different kinds of disclinations, and the red lines represent the contact force. 
  }
\label{schematic}
\end{figure}

The model system consists of $N$ identical elastic particles of mass $m_0$ confined in a two-dimensional
box of $L_0\times L_0$. The initial particle configuration is generated by random disk packing that
naturally creates inhomogeneity in density and stress
distribution~\cite{wan2018shapes}. In the initial random disk packing, the
particles of radius $R$ are not in contact. To promote efficiency of simulation, we increase
the radius of the particles in the initial state from $R$ to $R_0$. The volume fraction and stress
are thus controllable by specifying the radius $R_0$ of the expanded particles. A typical
random initial particle configuration is presented in Fig.~\ref{schematic}. In
our numerical experiment, the four walls of the box are simultaneously pushed
inward, and the side length becomes $L_0-\Delta L$ in the final state. And the
number density of the particles is $\rho=N/(L_0-\Delta L)^2$. In each
step of the compression process, $L$ is reduced by $\delta L$, which is set to
be sufficiently small to ensure that the system evolves quasi-statically.

The mechanical relaxation of the stressed system is governed by the particle-particle and
particle-wall interactions, both of which are modelled by the Hertzian
potential~\cite{landau99a}. Specifically,  the particle-particle interaction
potential is
\begin{displaymath} 
    V_{{\textrm {pp}}}(r)= \left\{ \begin{array}{ll} \frac{\sqrt{2R_0}}{5D}(2R_0-r)^{\frac{5}{2}} &
    r\leq 2R_0 \\ 
    0 &  r > 2R_0,
    \end{array} \right.  
\end{displaymath} 
where $r$ the distance between the particle centers, and $R_0$ is the
radius of the particle. $D = 3(1-\nu^2)/(2E)$, where $E$ is
the Young's modulus, and $\nu$ is the Poisson ratio. The interaction
potential between the particle and the rigid wall is 
\begin{displaymath} 
    V_{{\textrm {pw}}}(r)= \left\{ \begin{array}{ll} \frac{2\sqrt{R}}{5D'}(R_0-r)^{\frac{5}{2}} &
    r\leq R_0 \\ 
    0 &  r > R_0,
        \end{array} \right.  
\end{displaymath} 
where $r$ the distance from the center of the particle to the wall, and
$D'=D/2$.

\begin{figure}[t]  
\centering 
\includegraphics[width=3.4in]{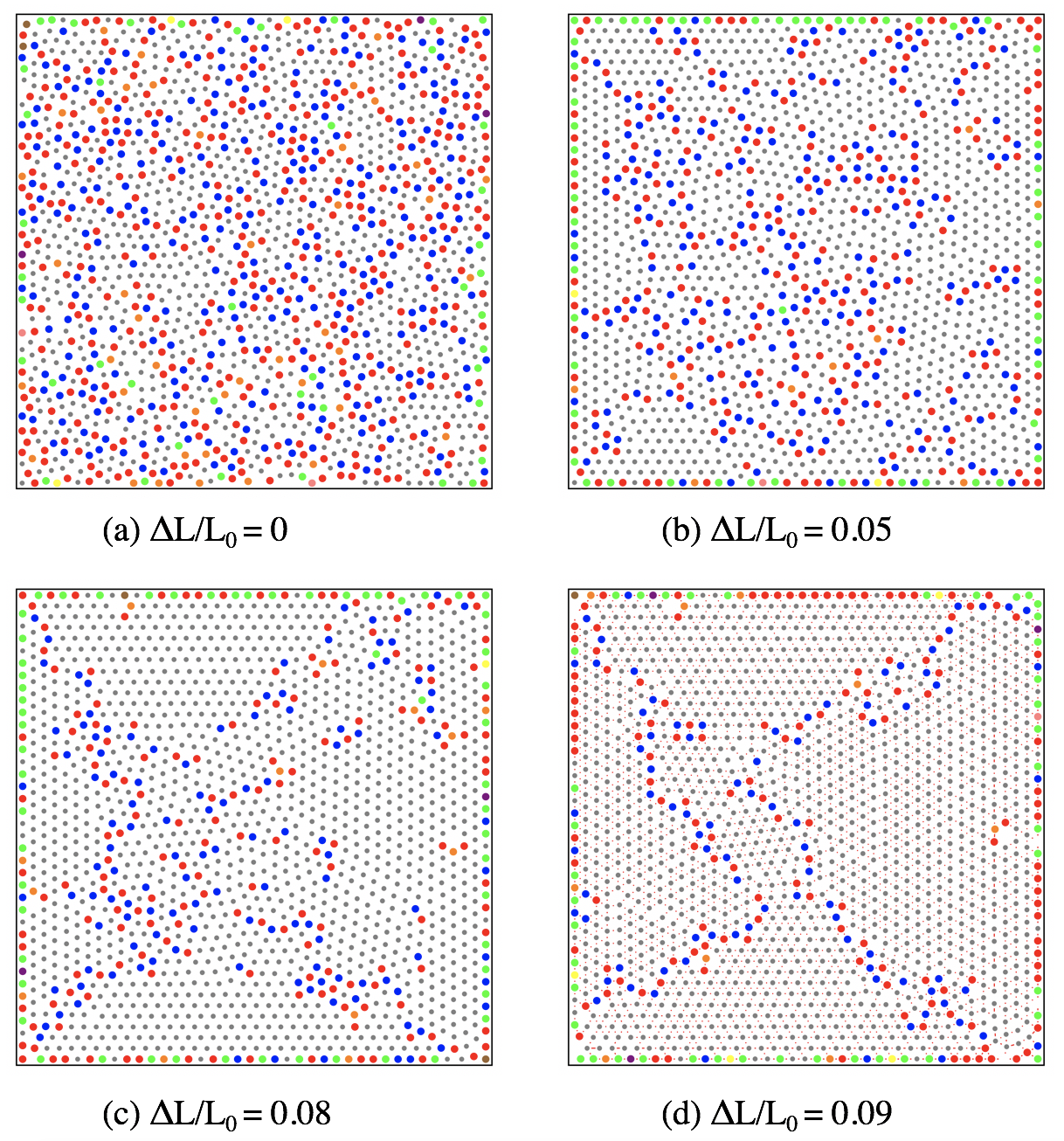} 
  \caption{Entropic ordering of random particle configuration into triangular
  lattice. All the configurations are in the lowest energy state.  Each dot
  represents a particle. The red and blue dots are the five- and seven-fold
  disclinations. The total energy $E=0$ (a)-(c), and $E= 1.12\times 10^{-6}$
  (d), which corresponds to a negligibly small overlap of particles. The
  compression rate $\delta L = 0.1$. $\rho=0.23$ (a), 0.25 (b), 0.26 (c), and
  0.27 (d). The mean lattice spacing is 2.25 (a), 2.14 (b), 2.07 (c), and 2.05 (d).
  $N=1600$. $L_0=84$. $R=0.7$.  }
\label{order}
\end{figure}

With a given box size, we perform the steepest descent method to find the lowest
energy state which corresponds to the optimal packing of the particles.
Specifically, the particle configuration is continuously updated by
simultaneously moving all of the particles along the direction of the force. The
displacement is proportional to the magnitude of the force. The typical step
size of the particle that is subject to the maximum force is $s=10^{-3} R_0$.
Evolution of the system is terminated when the slope of the energy curve becomes
sufficiently flat. Typically, termination occurs if the reduction rate of the
energy is less than $0.1\%$ in ten thousand steps.  In this work, we set
$R_0=1$, $m_0=1$, and $D=1$. The units of length, time, and mass are thus $R_0$,
$\sqrt{m_0 D/R_0}$, and $m_0$. By denoting $\tau_0=\sqrt{m_0 D/R_0}$, the unit
of force is $m_0 R_0/\tau_0^2$.

We resort to the combination of geometric, statistical and force analyses on the
lowest energy particle configurations. Specifically, by Delaunay triangulation,
one can identify different kinds of disclinations, which are elementary
topological defect in two-dimensional triangular
lattice~\cite{nelson2002defects}. An $n$-fold disclination refers to a vertex
whose coordination number $n$ is deviated from six. The crystallization process
can be well characterized by tracking the evolution of the emergent
disclinations. In the zoomed-in plot in Fig.~\ref{schematic}, we show the
five-fold (red dots) and seven-fold (blue dots) disclinations. The red lines
between the dots represent the contact force. The magnitude of the force is indicated by the
length of the line. We also analyze the diffraction pattern of the particle
configurations calculated from the structure function~\cite{Chaikin1995}: 
$S(\vec{q}) = \left< n(\vec{q})n(-\vec{q}) \right>$. The particle density 
$n(\vec{x})=\sum_{i=1}^N \delta(\vec{x}-\vec{x}_i)$, and $n(\vec{q})=\sum_{i=1}^N
\exp(-i\vec{q}\cdot \vec{x}_i)$.

\section{Results and discussion}

\begin{figure*}[th]  
\centering 
\includegraphics[width=7in]{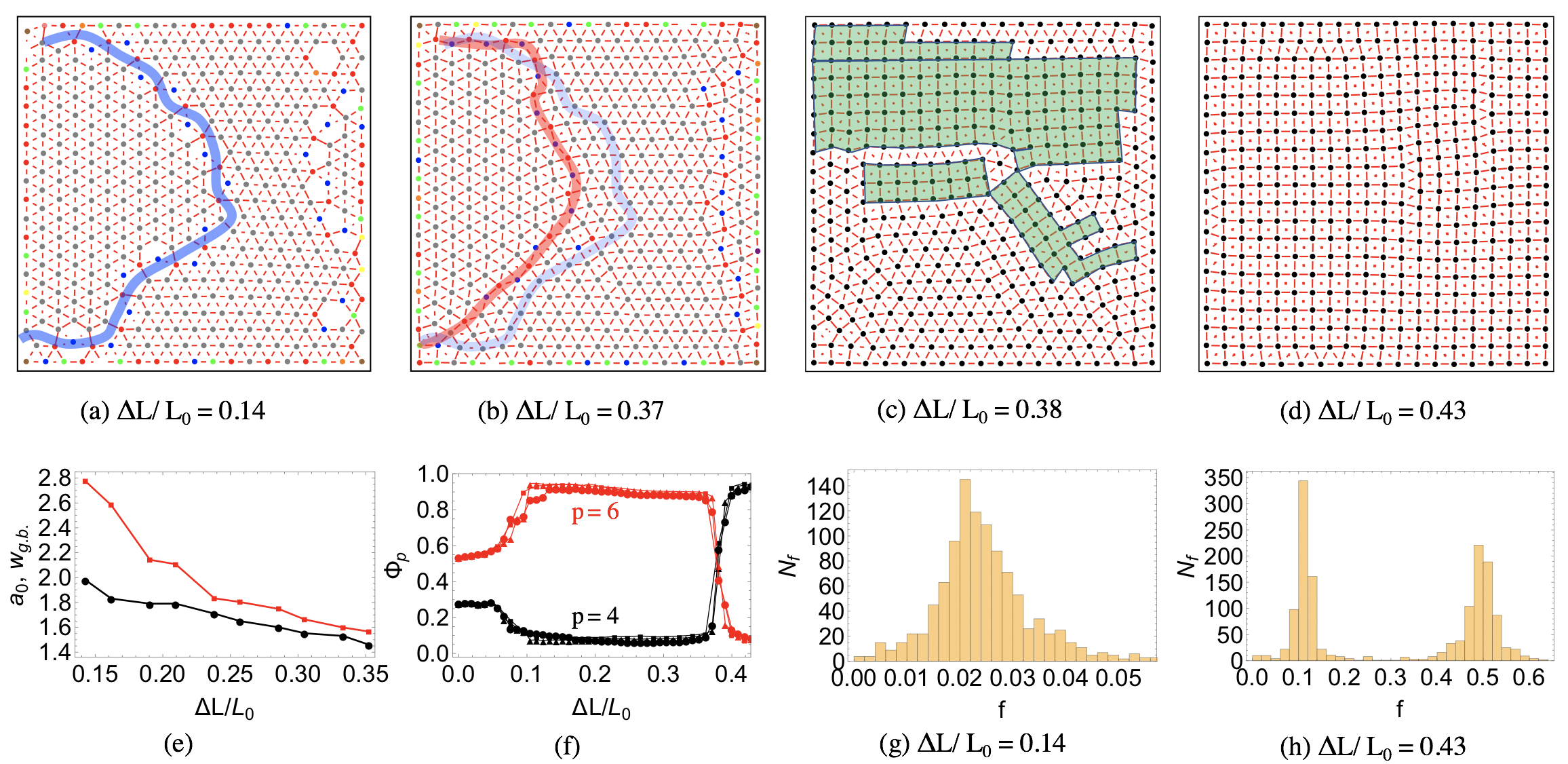}
  \caption{Triangular- to square-lattice transformation under compression.
  (a)-(d) Lowest energy particle configurations as the compression proceeds. In
  (a) and (b), the red and blue dots are the five- and seven-fold disclinations.
  The red lines represent the contact force.  The grain boundaries are indicated
  by the colored long curves. In (b), we also show the drift of the grain
  boundary with the compression of the system. In (c), the shadowed green
  domains indicate the square lattice in the background of the triangular
  lattice. $\rho=0.31$ (a), 0.57 (b), 0.58 (c), and 0.69 (d). The mean lattice
  spacing is 1.94 (a), 1.42 (b), 1.39 (c), and 1.29 (d). (e)-(h) Characterization of
  the lattice transformation.  (e) Faster reduction of the grain boundary width
  (red curve with squares) in comparison with that of the lattice spacing (black
  curve with circles). (f) Quantitative characterization of the
  bond-orientational order of the crystal lattice by the order parameter
  $\Phi_p$ (see main text for more information).  The three sets of red and
  black curves are obtained at the compression rates of $\delta L = 0.05$
  (circles), $\delta L = 0.1$ (triangles), and $\delta L = 0.2$ (squares),
  respectively.  (g) and (h) Splitting of the peak in the distribution profile
  of the contact force as the lattice transformation occurs. $N=400$.  $L_0=42$.
  $R=0.7$.  }
\label{tri_sq}
\end{figure*}

\begin{figure*}[t!h]  
\centering 
\includegraphics[width=6.6in]{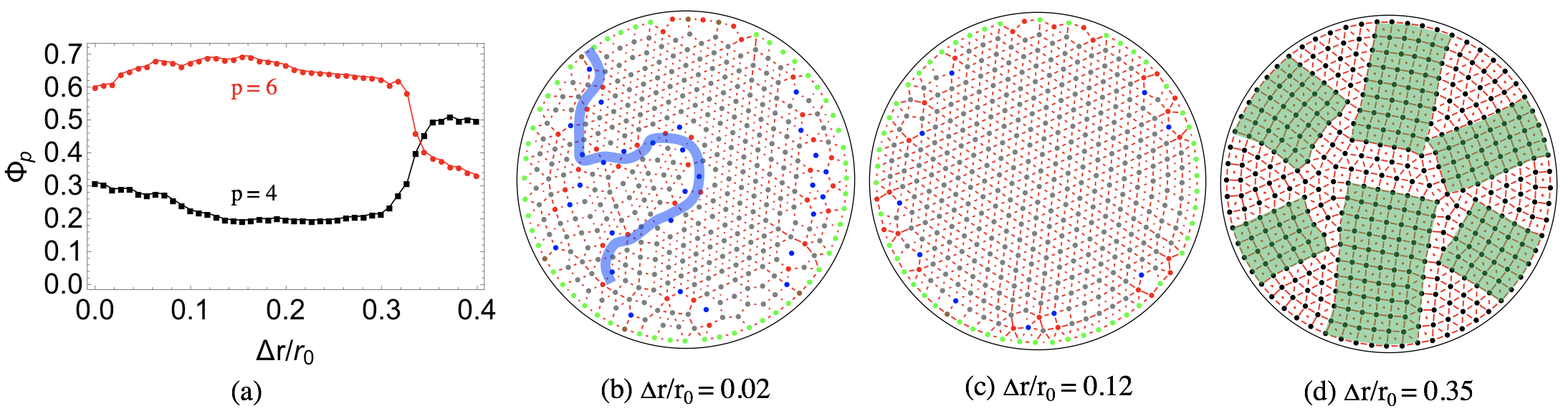}
  \caption{Triangular- to square-lattice transformation of a circular system
  under quasi-static compression. (a) Variation of the bond-orientational order
  in the lattice transformation (see main text for more information).  (b)-(d)
  Typical snapshots of lowest energy configurations as the compression proceeds.
  The colored dots represent different kinds of disclinations, and the red lines
  represent the contact force. $r_0$ is the initial radius of the
  circular boundary. $r_0=20.8$. (b) The grain boundary in the triangular lattice is
  indicated by the blue curve. (d) The square lattice is colored green. The
  compression rate $\delta r =0.1$. $\rho=0.96$ (b), 1.18 (c), and 2.17 (d). The
  mean lattice spacing is 1.94 (b), 1.75 (c), and 1.29 (d). $N=400$.}
\label{circular_4}
\end{figure*}

Figure~\ref{order} shows the growing region of triangular lattice as the system
is slightly compressed. The particles are represented by dots, and the
disclinations are indicated by different colors. The red and blue dots are the
five- and seven-fold disclinations. Figure~\ref{order}(a)-\ref{order}(c) show
the gradually established order in the arrangement of particles. Remarkably, the
particles are free of overlap in this process. It indicates that spherical
particles in a sufficiently crowded environment have the strong tendency of
spontaneously forming a triangular lattice. In Fig.~\ref{order}(d), the total
energy is at the order of $10^{-6}$, which corresponds to an average overlap of
particles as small as about $0.01\% R_0$. Therefore, the initial
crystallization of particles is driven by entropy instead of energy, and it is
known as entropic ordering~\cite{van2014understanding}. The microscopic dynamics
of this ordering process involves a series of defect events including
disclination annihilation and dislocation glide~\cite{nelson2002defects}.

Further compressing the system leads to the transition from triangular to square
lattice. Scrutiny of the typical particle configurations presented in
Fig.~\ref{tri_sq}(a) and \ref{tri_sq}(b) reveals the crucial role of the grain
boundary structure in resisting the increasing external compression and
preserving the triangular lattice. A grain boundary in a two-dimensional lattice
is a linear interface where crystallites of distinct orientations meet.
Figure~\ref{tri_sq}(a) shows that in the squeezed elastic particle system, the
grain boundary as indicated by the blue curve consists of a series of empty
pentagons. Under compression, these pentagonal voids
shrink much faster than the lattice spacing in the
crystallized regions. The quantitative result is presented in
Fig.~\ref{tri_sq}(e). It indicates that the
grain boundary structure plays a dominant role in absorbing the volume reduced
by the external compression. As the grain-boundary width shrinks, we also
observe the drift of the entire grain boundary towards the left boundary as
shown in Fig.~\ref{tri_sq}(b); the light blue curve indicates the relative
location of the grain boundary in Fig.~\ref{tri_sq}(a). Here, the drift of the
grain boundary structure is purely driven by external stress. Grain-boundary
migration as driven by the competition between thermal energy and the tendency to
minimize the grain-boundary curvature has been discussed in
Ref.~\cite{skinner2010grain-boundary}.

When the grain-boundary width is reduced to the size of the lattice spacing, a
slight compression from $\Delta L/L_0=0.37$ to $\Delta L/L_0=0.38$ triggers the
triangular-to-square lattice transition [see Fig.~\ref{tri_sq}(b) and
\ref{tri_sq}(c)]. The lattice transformation is initiated in a
few random domains that are colored in green in
Fig.~\ref{tri_sq}(c). Note that all the particle configurations in
Fig.~\ref{tri_sq} have reached the lowest energy state. The coexistence of the
square and triangular lattices in Fig.~\ref{tri_sq}(c) is a mechanically stable
state. Further compressing the system leads to the extension of the square
lattice to the entire system [see Fig.~\ref{tri_sq}(d)].

\begin{figure*}[th]  
\centering 
\includegraphics[width=7in]{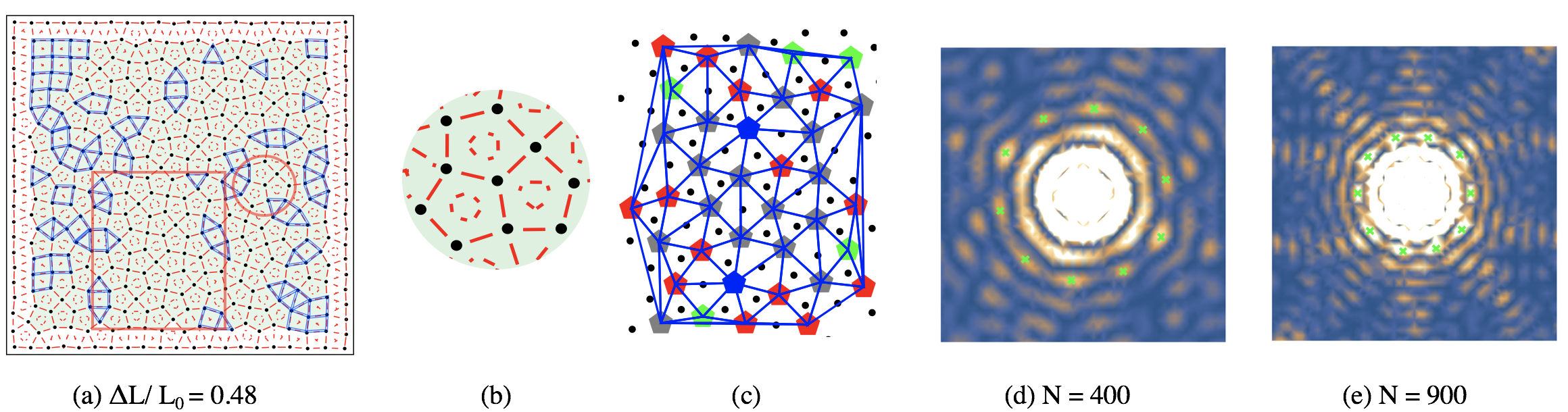}
  \caption{Formation of quasicrystal structure as the compression proceeds. (a)
  Tessellation of space by triangles, squares (both enclosed by blue lines), and
  nonregular pentagons. $\rho=0.84$. The mean lattice spacing is $1.2$. (b)
  Zoomed-in plot of the region in the red circle in (a). The red lines represent
  the contact force. (c) Delaunay triangulation of the pentagons in the red
  square box in (a).  The black dots represent the real particles.  (d) and (e)
  Diffraction patterns of the pentagonal tessellation in the systems of $N=400$
  and $N=900$ show the ten-fold rotational symmetry as indicated by the green
  cross marks.  $\delta L = 0.1$. $R=0.7$.  }
\label{pen}
\end{figure*}

To quantitatively characterize the triangular-to-square lattice transition, we
introduce the $p$-fold bond-orientational order parameter $\Phi_p$ defined
as~\cite{nelson2002defects}
\begin{eqnarray}
  \Phi_p = \frac{1}{N} \sum_{i=1}^{N} | \frac{1}{n_b} \sum_{j=1}^{n_b}
  \exp(ip\theta_{ij}) |,
\end{eqnarray} 
where $n_b$ is the number of
neighbors surrounding the particle $i$, $\theta_{ij}$ is the angle between the line connecting the
particles $i$ and $j$ and some chosen reference line, $N$ is the total number of
particles. The bond order in triangular and square lattices can be characterized
by $\Phi_4$ and $\Phi_6$, respectively. The
dependence of $\Phi_4$ and $\Phi_6$ on $\Delta L/L_0$ is presented in
Fig.~\ref{tri_sq}(f). With the increase of compression, 
the six-fold bond-orientational order emerges when $\Delta L/L_0 >0.1$, and the
triangular-to-square lattice transition occurs at $\Delta L/L_0 \approx 0.4$.
Simulations of systems with $N$ ranging from $100$ to $2000$ show that the
critical values for the appearance of triangular lattice and the lattice
transformation are independent of the number of particles. A further inquiry is if the formation
of the square lattice is due to the square shape of the boundary. To address
this question, we simulate systems with circular boundary, and also find
triangular-to-square lattice transition, as well as the coexistence of
triangular and square lattices as a mechanically stable configuration.  The
critical value for the triangular-to-square transition in the circular system is
identical to that of the
square system, as presented in Fig.~\ref{circular_4}.

In the triangular-to-square lattice transition, it is found that the
distribution of contact force $f$ is qualitatively changed. $f$ is the magnitude
of the force between two particles in contact. Statistical
distributions of the magnitude of the contact force are plotted in
Fig.~\ref{tri_sq}(g) and \ref{tri_sq}(h) for the triangular and square lattices,
respectively. We see that the lattice transformation leads to the split of the
peak structure in Fig.~\ref{tri_sq}(g).  The separation of the contact force
distribution is also seen in Fig.~\ref{tri_sq}(b) and \ref{tri_sq}(d), where the
length of the red lines represent the magnitude of the contact force. The short
lines at the center of each square element in Fig.~\ref{tri_sq}(d) correspond to
the peak at the lower end of the force distribution in Fig.~\ref{tri_sq}(h).

Here, we shall mention that the fabrication of square lattice generally involves
anisotropic interactions. Various sophisticated schemes have been proposed to
create anisotropic interactions, including design of particle shape and
chemistry~\cite{knorowski2012dynamics, li2013topological,girard2019particle}, application of electro-magnetic
fields~\cite{biswal2004rotational} and capillary
interactions~\cite{ershov2013capillarity, liu2018capillary}. In our system, the
triangular- to square-lattice transformation is driven by the microscopic
elastic instability of the squeezed particles under enhanced stress. This
elasticity-based scheme may provide a general strategy in the broad contexts of
soft matter packing like directed self-assembly and colloidal crystallization.

The triangular-to-square lattice transition can be understood by the following
energetic calculation. Consider a collection of $N$ particles in triangular and
square lattices in a square box of side length $L$, respectively. Geometric argument
shows that the total energies of these two systems are: $E_3=3N\epsilon_3$, and
$E_4=2N\epsilon_4$. $\epsilon_3$ and $\epsilon_4$ are the energy costs for a pair
of particles in contact in the triangular and square lattices. When $L<L_c$, $E_4(L)$ becomes smaller than $E_3(L)$; 
the system is thus dominated by the energetically favored square lattice. The critical value $L_c$
is derived as:
\begin{eqnarray}
  L_c=2 c_0 \sqrt{N} R_0,
\end{eqnarray}
where $c_0= ((3/2)^{2/5}-1)/(108^{1/20}-1)\approx 0.67$, and $R_0$ is the radius
of the particle. For $N=400$, we have
$L_c \approx 26.7$. Simulations at varying compression rate (from $\delta L
=0.05R_0$ to $\delta L =0.2R_0$ in each step) and initial particle radius $R$
(from $R=0.5$ to $R=0.7$) return a common critical value of $L_c \approx 26.0$.
These two critical values from numerical experiment and theoretical analysis are
very close, and it shall be related to the friction-free feature of the
particles that can facilitate the lattice transformation.

\begin{figure*}[t!h]  
\centering 
\includegraphics[width=6.6in]{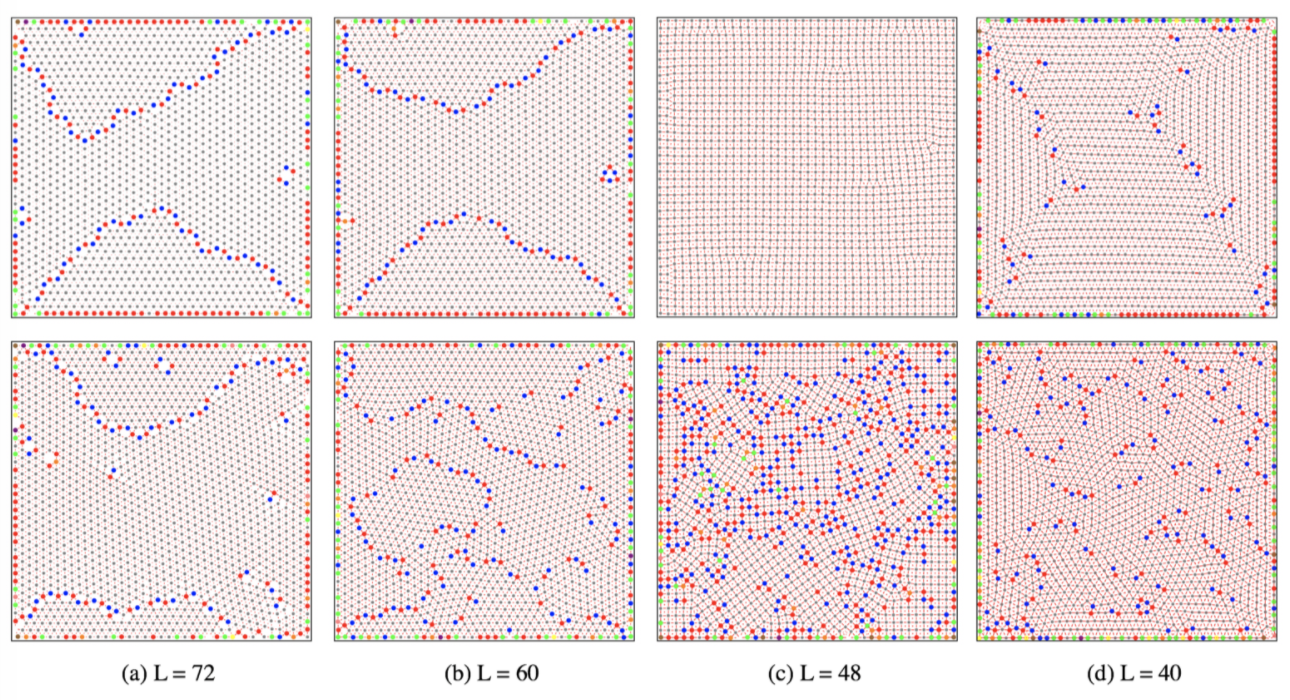}
  \caption{Imposing compression significantly improves ordered packing of
  elastic particles, especially in the regime of high particle density.  Upper
  panels: the evolution is driven by a single gradual compression process. The system is initially free
  of stress at $L=L_0=84$. $\delta L=0.1$. Lower panels: lowest energy
  configurations via relaxation of randomly distributed particles in the boxes of
  given sizes. The colored dots represent different kinds of disclinations, and the
  red lines represent the contact force.  $N=1600$.   }
\label{random_order}
\end{figure*}

\begin{figure}[t!h]  
\centering 
\includegraphics[width=3.6in]{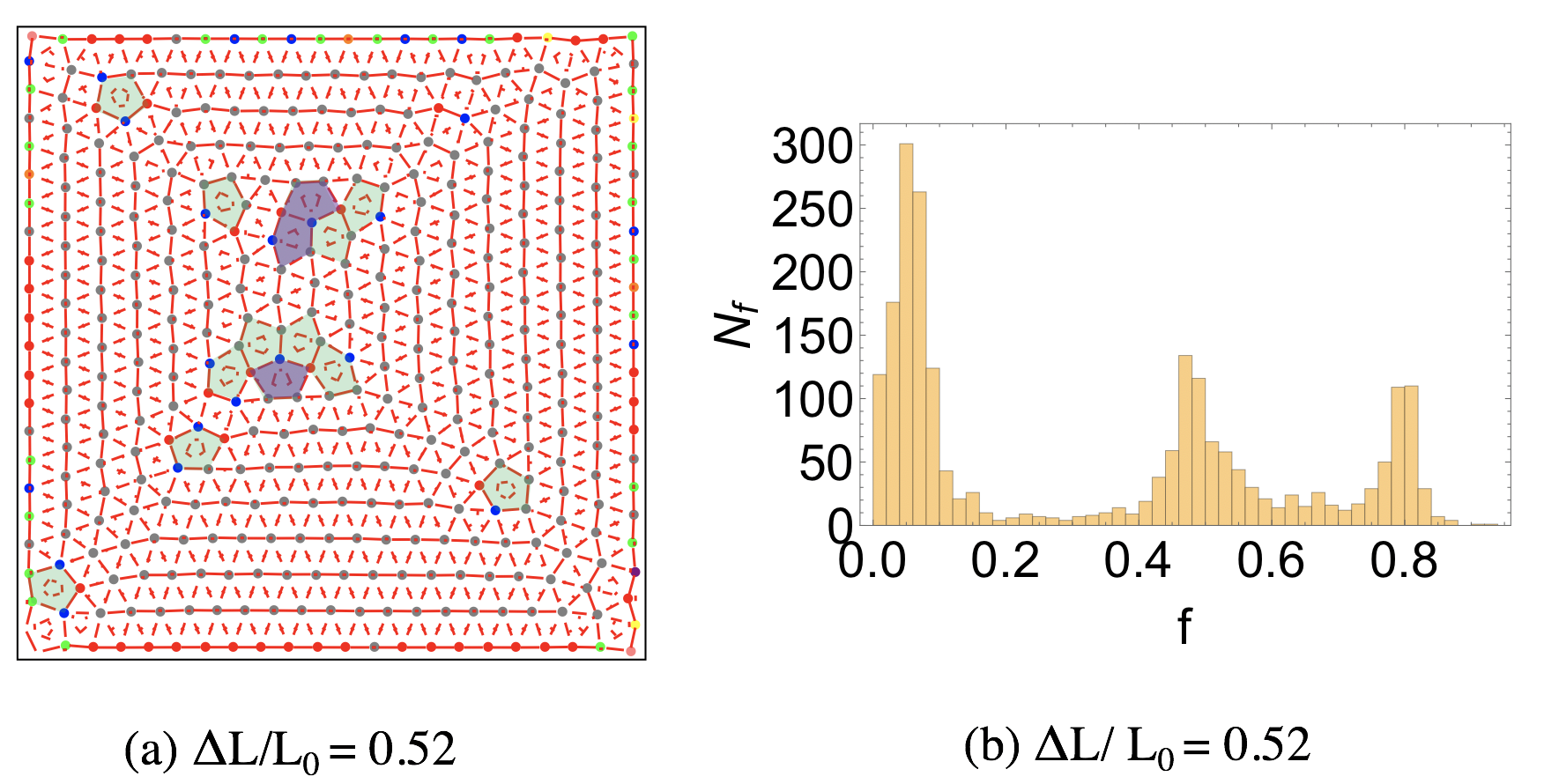}
  \caption{Reentrant configurational transition to triangular lattice at larger
  compression.  (a) The triangular lattice with scattered dislocational (green)
  and disclinational (purple) pentagonal vacancies. The red lines represent the
  contact force. (b) Further splitting of the peak in the distribution profile
  of the contact force as the lattice transformation occurs [in comparison with
  Fig.~\ref{tri_sq}(g) and \ref{tri_sq}(h)]. $\rho=0.98$. The mean lattice
  spacing is $1.1$. $N=400$.  $\delta L = 0.1$.
  $L_0=42$. $R=0.7$.}
\label{tri}
\end{figure}

As the compression proceeds, the square lattice is transformed to a pattern as
shown in Fig.~\ref{pen}(a). We observe the tessellation of space by the three
kinds of elementary shapes: squares, triangles, and pentagons. The zoomed-in
plot in Fig.~\ref{pen}(b) shows that each pentagon is characterized by five long
red lines connecting the five particles and five short lines inside the
pentagon.  The short lines represent the contact of a particle with its next
nearest neighbors around the pentagon. We emphasize that the introduction of
contact force lines allows us to unambiguously identify the pentagonal
units.  Comparison of the straight
contact force lines in the rectangular lattice in
Fig.~\ref{tri_sq}(d) and the zig-zag force lines in Fig.~\ref{pen}(a) shows that
the pentagonal structures originate from the local elastic instability
characterized by the buckling of the contact force lines. Figure~\ref{pen}(c)
shows the packing of the emergent pentagonal units. The black dots represent
real particles. Each pentad of particles is represented by a pentagon.
Delaunay triangulation reveals the underlying topological structure. The red and
blue pentagons indicate the five- and seven-fold disclinations in the
triangulated pentagons. These pentagonal objects enrich the scenario of regular
packing and 2D crystallography~\cite{bowick2002crystalline,
Bausch2003e,miller2011two,guerra2018freezing}.

The tessellation of a continuous flat region by pentagons is
remarkable. As a geometric restriction, regular pentagonal tiling on the
Euclidean plane is impossible. Regular pentagons can tile the hyperbolic plane
and the sphere though~\cite{conway2016symmetries}. Here, the key to achieving the
pentagonal tessellation is the deformability of the pentagons due
to the softness of the particles~\cite{bowers1997regular}. The tessellation of deformable pentagons
implies an induced non-Euclidean geometric structure in the particle array
under high compression~\cite{struik88a, yao2013topological, soni2018emergent}.
The nonregular pentagonal tessellation realized in soft particle arrays
explains the extensive existence of quasicrystals in soft matter
systems~\cite{jagla1998phase, miller2011two, dotera2011quasicrystals,
dotera2014mosaic, zu2017forming}. In our system, we reveal
the ten-fold rotational symmetry in the diffraction patterns as a characteristic
of a quasicrystal, which is indicated by the ten green cross marks in
Fig.~\ref{pen}(d) and \ref{pen}(e). It is in contrast with the eight- and
twelve-fold rotational symmetries of quasicrystals found in unconstrained
soft particle systems in Ref.~\cite{zu2017forming}.

The distinct rotational symmetries of the quasicrystal structures reported in
Ref.~\cite{zu2017forming} and that in our work suggest the subtle kinetic effect
of external stress on the packing of soft particles, which is beyond merely
changing the particle density. We notice that in our system crystallization is
initiated at the boundary as the compression proceeds. To further illustrate the
kinetic effect of the external stress, we examine the mechanical relaxation of
randomly distributed particles in boxes of given sizes, and compare the
resulting lowest energy configurations with those obtained via the compression
procedure. In Fig.~\ref{random_order}, the upper panels show the lowest energy
configurations in a single gradual compression process. The lower panels show
the corresponding lowest energy configurations upon the same amount of
compression, but each of them is via the independent relaxation of random
particle distribution. Comparison of these two kinds of cases clearly shows that
imposing compression can significantly facilitate ordered packing, especially
for the systems with relatively high particle density.

Under even larger compression, it is observed that the pentagonal tessellation
becomes unstable, and the system returns to the triangular lattice, as shown in
Fig.~\ref{tri}. This reentrant behavior can be attributed to the effect of
density variation under compression~\cite{narayanan1994reentrant,
miller2011two}. Scattered dislocational (green) and disclinational (purple)
pentagonal vacancies are observed in the triangular lattice.  Pentagonal
vacancies of distinct topological nature have been found in Lennard-Jones
crystals confined on the sphere in our previous work~\cite{yao2017topological}.
Statistical distribution of the contact force in the configuration of
Fig.~\ref{tri}(a) is presented in Fig.~\ref{tri}(b). The number of peaks
increases from two [see Fig.~\ref{tri_sq}(h)] to three. Furthermore, the peaks
experience rightward drift under enhanced compression. The close connection of
contact force distribution and particle configuration suggests the contact-force
analysis as a useful tool for revealing important structure information in
deformable particle systems. Here, we shall note that high compression may be
realized in a series of deformable mesoparticle systems, such as ultrasoft colloids
and hydrogel beads~\cite{Likos2006, miller2011two}. The Hertzian model, which is
based on the linear continuum elasticity theory, may not quantitatively describe
the relevant soft interactions~\cite{brodu2015, pamies2009phase, miller2011two}.
The strong repulsion between two close particles is qualitatively
captured by the Hertzian potential. As such, the numerical results in
Figs.~\ref{pen} and \ref{tri} present a plausible scenario of ordered packing of
elastic particles under high compression.

\begin{figure}[t!h]  
\centering 
\includegraphics[width=2in]{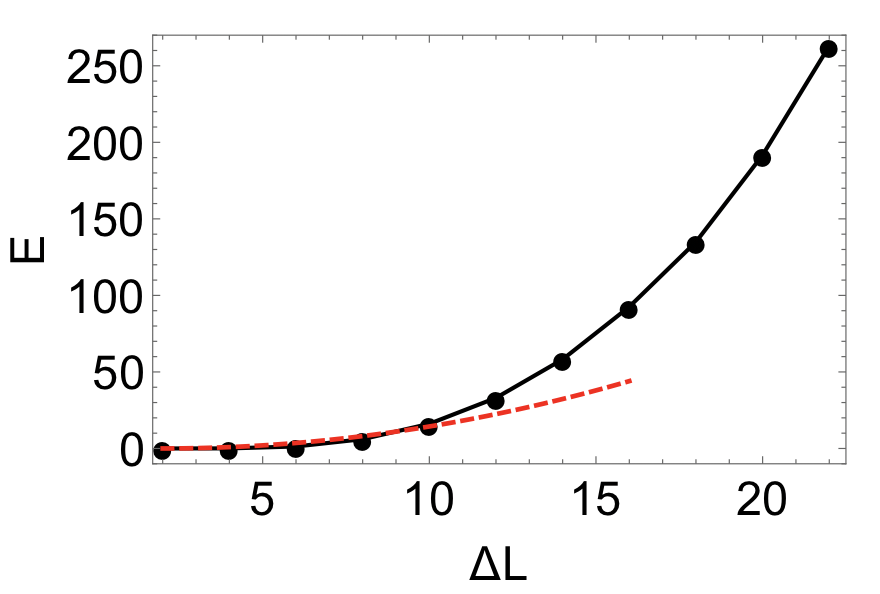}
  \caption{Plot of total energy versus the amount of compression $\Delta L$. $E$ is the energy of
  the relaxed particle configuration at a given box size $L=L_0-\Delta L$. $L_0=42$. The red
  dashed curve is a quadratic fitting. $\delta L=0.1$. $N=400$.
   $R=0.7$.   }
\label{energy}
\end{figure}

We finally briefly discuss the robustness of the compression-driven ordered
packing, the reversibility issue and role of dimension on particle packing. In
our preceding discussion, we show that external compression eliminates the
randomness in the initial particle configuration, and pushes the system to
develop a series of ordered structures from triangular and square lattices [see
Fig.~\ref{tri_sq}] to pentagonal
tessellation [see Fig.~\ref{pen}], and finally back to the triangular lattice
[see Fig.~\ref{tri}]. In this process, as shown in Fig.~\ref{energy}, the system
energy increases with $\Delta L$ at a faster rate than the quadratic law, which
indicates an enhanced rigidity of the system. Systematic simulations at varying
compression rate ($\delta L$ ranges from $0.05$ to $0.2$) and $R$-controlled
initial stress level ($R$ ranges from $0.5$ to $0.8$) reveal that the particle
configuration always evolves by the same sequence. By reversing the compression
process, could the system exhibit any memory effect? To address this question,
we start from the triangular lattice under high compression, and gradually
enlarge the box size.  It turns out that the sequence of order is revered. At
the end of the cycle of compression and expansion, the initial randomness in the
packing of particles is eliminated, and the system terminates at the stress-free
triangular lattice. Our simulations imply that two-dimensional monodisperse
elastic particles have the natural tendency to form ordered phases.  This
tendency can be attributed to the lack of geometrical frustration; the local
dense packing is compatible with the global dense packing in the form of triangular
lattice~\cite{atkinson2014existence}. In contrast, a three-dimensional system is
intrinsically geometrically frustrated, because the preferred local icosahedral
structure cannot be extended to the whole flat
space~\cite{nelson1989polytetrahedral, atkinson2014existence}.

\section{Conclusion}

In summary, by the Hertzian particle model, we have illustrated the crucial role
of stress in organizing particles from random to ordered states in 2D packings. Specifically,
we have revealed regular packings from triangular and square lattices to the
remarkable quasicrystal structure, and to the reentrant triangular lattice under
increasing compression. We have also identified the grain boundary structure as
an energy absorber to stabilize the triangular lattice, and characterized the
regular packing at high compression in terms of pentagonal tessellation. The
physical origin of ordered packing is essentially from the interplay of external
stress and intrinsic elastic instability of the particle array. In combination
with the broad possibility of interaction potential design in soft matter
systems, exploiting the role of stress as an ordering field may provide a new
dimension for realizing ordered packings in various 2D cases like self-assembly
and colloidal crystallization on interfaces. Due to the intrinsic geometric
frustration in 3D systems that is absent in planar 2D systems, it is of interest
to further explore the role of stress in regulating 3D packings that may lead to
new physics not seen in 2D systems.

\section{ACKNOWLEDGMENTS}

This work was supported by NSFC Grants No. 16Z103010253. The author thanks the
support from the Student Innovation Center at Shanghai Jiao Tong University.

\end{document}